\documentclass[%
 reprint,
 amsmath,amssymb,
 aps,
]{revtex4-2}

\usepackage{color}
\usepackage{hhline}
\usepackage{soul}
\usepackage{xfrac}
\usepackage[normalem]{ulem}
\usepackage{float}
\usepackage{subcaption}

\usepackage{graphicx}
\usepackage{dcolumn}
\usepackage{bm}
\usepackage[mathlines]{lineno}

\usepackage[utf8]{inputenc}
\usepackage[T1]{fontenc}
\usepackage{mathptmx}
\usepackage{amsmath}
\usepackage{amssymb}
\DeclareCaptionJustification{justified}{\leftskip=0pt \rightskip=0pt \parfillskip=0pt plus 1fil}

\draft

\begin{document}

\preprint{APS/123-QED}

\title{Estimating the Steady State Diffusion Coefficient 
\\Using Data from the Transient Anomalous Regime} 

\author{Nicholas Ilow}
\email[]{nilow051@uottawa.ca}
\affiliation{University of Ottawa, 75 Laurier Ave. E, Ottawa, ON K1N 6N5}

\author{Gary W. Slater}
\email[]{gary.slater@uottawa.ca}
\affiliation{University of Ottawa, 75 Laurier Ave. E, Ottawa, ON K1N 6N5}

\date{\today}

\begin{abstract}
When particles/molecules diffuse in systems that contain obstacles, the steady-state regime (during which the mean-square displacement scales linearly with time, $\left< r^2 \right> \sim t$) is preceded by a transient regime. It is common to characterize this transient regime using the concept of anomalous (sub)diffusion with the scaling law $\left< r^2 \right> \sim t^\alpha$, where the corresponding exponent $\alpha<1$. We propose a new method to estimate the critical time $t^*$ that marks the transition between these two regimes. The method uses short-time data from the transient regime to estimate $t^*$, which can then be used to estimate the steady-state diffusion coefficient $D$. In other words, we propose a procedure that makes it possible to estimate the steady state diffusion coefficient without reaching the steady-state. We test the procedure with various two-dimensional lattice systems.
\end{abstract}


\maketitle 

\section{Introduction}

Diffusion in the presence of obstacles, often called obstructed diffusion, is present in a wide range of physical, chemical and biological systems \cite{ratto2002obstructed,rogers2013obstructed,weigel2012obstructed,shorten2009mathematical,anderson2021subtle}. Generally speaking, we can expect two different time regimes. At short times, the diffusing particles explore the spatial constraints and start colliding with the obstacles \cite{metzler2014anomalous,krapf2015mechanisms} (this is sometimes preceded by a free diffusion regime if the obstacles are far from one another \cite{hofling2013anomalous,stolle2019anomalous}). The steady-state, which is reached at long times, is characterized by the fact that the mean squared displacement (MSD) of a tracer particle is linear in time, with some excess contribution (to be denoted $\beta^2$) which is due to the fact that diffusion is faster at early times \cite{hofling2013anomalous} (the collisions slow down the diffusion process):
\begin{equation}
    \label{eq:GeneralSecondMoment}
    \left< r^2(t) \right> = \beta^2 + 4Dt~,
\end{equation} 
where $D$ is the diffusion coefficient. Therefore, in order to access the diffusion coefficient, the experimental or simulation time must be much larger than the crossover time $t^*$ between the transient and steady-state regimes\cite{Random_Obstacles_Saxton}.

Equation~\ref{eq:GeneralSecondMoment} is not valid during the transient regime. Instead, the MSD is often fitted using the concept that it will follow a power law with an "anomalous" exponent $\alpha$ \cite{metzler2000random,sokolov2012models}
\begin{equation}
    \label{eq:TransientSecondMoment}
    \left< r^2(t) \right> = 4D_\alpha t^\alpha~,
\end{equation}
where in our case $\alpha<1$. Consequently, this regime is also called the anomalous diffusion regime. In practical cases, defining the time range over which eq.~\ref{eq:TransientSecondMoment} might be valid is often arbitrary \cite{ArbitraryAlpha7,alcazar2018general,ellery2016analytical,ArbitraryAlpha,ArbitraryAlpha3}. In Sections~\ref{sec:data-analysis} and \ref{sec:tStarEstimate}, we propose methods to determine the center $t_I$ and the width $\Sigma_t$ of the anomalous regime, respectively.

Using eqs.~\ref{eq:GeneralSecondMoment} (with $\beta=0$) and \ref{eq:TransientSecondMoment}, we can estimate the crossover time $t^*$. The time at which both equations predict the same displacement is simply
\begin{equation}
    \label{eq:t*}
    t^* = \left({D_\alpha}/{D}\right)^{1/({1-\alpha})},
\end{equation}
while the corresponding crossover distance $r^*$ is given by
\begin{equation}
    \label{eq:r*}
    r^{*2} = 4D \, t^* = 4 D_\alpha \, t^{*\alpha} = {\left({D_\alpha}/{D^\alpha}\right)^{{1}/(1-\alpha)}}.
\end{equation}
Note that obtaining $t^*$ and $r^*$ requires knowledge of the steady-state diffusion coefficient $D$. 

The main result of this paper can be summarized as follows: the width $\Sigma_t$ of the anomalous regime, as we define it in Section~\ref{sec:tStarEstimate}, is an excellent approximation for the crossover time $t^*$, even though it does not require any knowledge of the diffusion coefficient $D$. This has an important and direct application since we can rearrange eq.~\ref{eq:r*} as
\begin{equation} \label{eq:D_Estimate_tStar}
    D = D_\alpha \, t^{*^{\alpha-1}}~.
\end{equation} 
This equation implies that if we could find a proxy for $t^*$ using data from the anomalous regime, we would be able to estimate the steady state diffusion coefficient $D$ without reaching the steady state. As we shall demonstrate, this is precisely what our definition of $\Sigma_t$ allows us to do.

Experimentally diffusion problems can be explored using a range of techniques including Fluorescence Recovery After Photobleaching \cite{arrio2000translational, seksek1997translational}, and Single Particle Tracking \cite{murase2004ultrafine, kusumi2005single}. However, most methods do not easily allow experimentalists to probe both regimes \cite{MONNIER2012616,10.3389/fphy.2020.00134} because of the wide range of time scales involved. Our paper thus offers a potential approach to estimating the steady-state diffusion coefficient $D$ using short-time data only.

In this manuscript we introduce (Sections~\ref{sec:data-analysis}--\ref{sec:tStarEstimate}) and test (\ref{sec:testing}) our new concepts using a simple model of obstructed diffusion with random obstacle configurations on a two dimensional square lattice (Section~\ref{Methods}) since this allows us to obtain high-precision numerical data for both the anomalous and steady-state regimes.

\section{Methodology} \label{Methods}

We consider random walks on a square lattice with a mesh size $a$. The standard Lattice Monte Carlo (LMC) algorithm \cite{weiss1994aspects,hughes1998random} includes random jumps of length $a$ along one of the four Cartesian directions, each with a probability $p=\tfrac{1}{4}$. The second moment of the distribution of displacements after $N$ steps is then given by $\left<r^2\right>_G = Na^2$; if the duration of these MC steps is $\tau$, this can be rewritten as $\left<r^2\right>_G = 4Dt$, with $D=a^2/4 \tau$ the diffusion coefficient and $t=N \tau$ the time. Since the second moment is identical to the one predicted by the solution of the diffusion equation (which is a Gaussian distribution, hence the subscript G), LMC models are often used to simulate diffusion problems. However the 2D LMC algorithm actually gives the following fourth moment
\begin{equation}
\label{eq:r4}
    \left<r^4(t)\right> = 32D^2t^2 + (4-32p)Dta^2~,
\end{equation} 
while the solution of the diffusion equation gives $\left<r^4(t)\right>_G = 32D^2t^2$. The last term in eq.~\ref{eq:r4} is negligible at long times, but it does affect the mean time between collisions with obstacles and hence impacts the transient regime including the value of the anomalous exponent $\alpha$ (data not shown). To eliminate this correction term and improve the algorithm, we use $p=\frac{1}{8}$ and a probability $p_o=1-4p=\tfrac{1}{2}$ of staying put during a time step. 

Typically, obtaining the steady-state diffusion constant $D$ would require long LMC simulations. However calculating $D$ for a random-walker in a particular system of randomly distributed obstacles (which act as reflecting boundaries and occupy a fraction $\phi$ of the surface area) with periodic boundary conditions is made possible using the numerical methods outlined in \cite{Mercier}. This approach first involves computing the steady state concentration profile under a weak bias $\epsilon$. This is done by solving an $N\times N$ matrix, where $N$ is the number of lattice sites accessible to the particle, and each row is determined by a rate equation with biased jumping probabilities. One can then calculate the mean velocity using the computed concentration profile and the local velocities resulting from the bias and the presence of reflecting obstacles. Finally, the mean velocity $v(\epsilon)$ is used to compute the diffusion constant $D$ via the Nernst-Einstein relation:
\begin{equation}
\label{eq:Nernst}
    \frac{D(\phi)}{D(0)} = \lim_{\epsilon\to0}~ \frac{v(\epsilon,\phi)}{v(\epsilon,0)}~.
\end{equation}
We use this approach to obtain very high precision estimates of $D$ below. For these calculations, we use a lattice of size $L = 128$ and an ensemble size of $2000$. A lattice spacing of $a=1$ is chosen, with $\tau=\frac{1}{8}$ corresponding to the required timestep for $p=\frac{1}{8}$, such that $D(0) = 1$.

We use a Markov Chain Monte Carlo (MCMC) method to obtain high-precision short-time data. Since the obstacles are reflecting boundary conditions, the concentration evolves (starting with a unit concentration on a single site at the center of the lattice) as follows: 
\begin{equation}
    \label{eq:ConcentrationUpdate}
    \begin{aligned}
        C_{x,y}(t+1) &= [C_{x-1,y}(t) + C_{x,y}(t)O_{x+1,y}]p_{+x}\\
        &+[C_{x+1,y}(t) + C_{x,y}(t)O_{x+1,y}]p_{-x}\\
        &+ [C_{x,y-1}(t) + C_{x,y}(t)O_{x,y+1}]p_{+y}\\
        &+ [C_{x,y+1}(t) + C_{x,y}(t)O_{x,y-1}]p_{-y}\\
        &+ [C_{x,y}(t)]p_{o},
    \end{aligned}
\end{equation}
where $C_{x,y}(t)$ is the concentration at lattice site $(x,y)$ at time (or iteration) $t$, and $O_{x,y}$ is a binary value ($0$ or $1$) describing the presence of an obstacle at $(x,y)$. Note that $p_{\pm x} = p_{\pm y}=p=\frac{1}{8}$. The MCMC calculations are completed on a lattice of size $512 \times 512$, and we handle disorder by averaging over an ensemble of $2000$ different obstacle configurations.

A feature of random systems is the possible presence of closed areas, which we call "lakes". A lake is an area in which unoccupied sites exist, but these sites are inaccessible to a tracer particle initially located outside. We must make sure that a Markov chain calculation does not start in a lake since $\left<r^2(t)\right>$ would then quickly plateau. When we calculate $D$ using exact matrix calculations and eq.~\ref{eq:Nernst}, lakes lead to block-diagonal matrices, and the only non-zero value of $D$ is found for the block that corresponds to the connected pathway through the network. To avoid such issues, we simply fill all unoccupied sites within lakes with phantom obstacles. 

\section{Results}

\subsection{Data analysis}
\label{sec:data-analysis}

\begin{figure}[ht]
    \centering
    \includegraphics[scale=1]{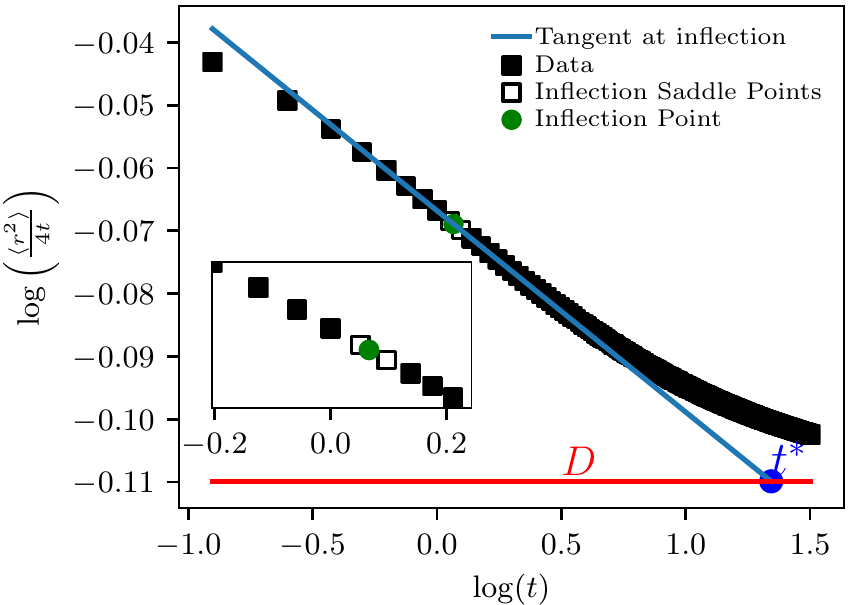}
    \caption{Log-log plot of the instantaneous diffusivity ratio $ {\left<r^2\right>}/{4t}  $ \textit{vs.} time $t$ for a $1\times1$ random-walker, with $1\times1$ obstacles at an obstacle concentration $\phi = 10\%$. The steady-state diffusion coefficient $D$ (horizontal red line) was calculated using the matrix technique described in Section~\ref{Methods}, and the slope of the line tangent to the inflection point is $\alpha -1$. Where these two lines intersect defines the crossover time $t^*$. }
    \label{fig:logDlogT}
\end{figure}

Central to our methodology is plotting $\log \, \langle r(t)^2\rangle / 4t$ \textit{vs} $\log(t)$. As Fig.~\ref{fig:logDlogT} shows, the transient and steady-state regimes are clearly visible when the data are plotted in this way. Since the location and boundaries of the transient regime are often defined arbitrarily \cite{ArbitraryAlpha7,alcazar2018general,ellery2016analytical,ArbitraryAlpha,ArbitraryAlpha3}, the value of the anomalous exponent $\alpha$ is generally method-dependent. However, an inflection point is present for all cases in this type of log-log plot. We thus propose that the inflection point provides a robust way of defining the center of the anomalous regime. The anomalous exponent can then be extracted from the slope of the tangent at the inflection point (which is $\alpha - 1$ given eq.~\ref{eq:TransientSecondMoment}) while $D_\alpha$ is simply given by the ratio $r_I^{2}/4t_I^{\alpha}$ (the subscript $I$ refers to evaluation at the inflection point). We investigate the width of the anomalous regime in Sec.~\ref{sec:tStarEstimate}. Figure~\ref{fig:logDlogT} further defines the crossover time $t^*$ as the intersection of the tangent and the horizontal line marking the steady state diffusion coefficient $D$. The inset details the numerically calculated inflection saddle points, thus determining the fitting region for estimating the interpolated position of the inflection point and both $\alpha$ and $D_\alpha$. 

We present the results of $t^*$, $r^{*} = \sqrt{4Dt^*}$ and $\alpha$, for a $1 \times 1$ walker in random systems with $1 \times 1$ obstacles for varying obstacle concentrations $\phi$ in Fig.~\ref{fig:No_D}. At larger obstacle concentrations the transition length scale $r^*$ grows to values such that our system size ($L = 512$) is no longer large enough to avoid finite-size effects ($r^*$ is related to the crossover length or cluster size in percolation theory \cite{Percolation}). The transition time $t^*$ behaves similarly. The exponent $\alpha$, on the other hand, decreases continuously with $\phi$ to reach a value of about $\alpha=0.733$ at $\phi=\phi^*$. Finally, the blue data points in Fig.~\ref{fig:D} show how the diffusion coefficient $D$ decays with $\phi$; in particular, we see that $D \to 0$ at the percolation threshold $\phi = \phi^* \approx 0.40725$ (this actually defines $\phi^*$).

\subsection{The width of the anomalous regime}
\label{sec:tStarEstimate}

Obviously, the transient regime can only be said to satisfy eq.~\ref{eq:TransientSecondMoment} over a small region centered around the inflection point in Fig.~\ref{fig:logDlogT}. Since the slope at this point is $\alpha-1$ while the second derivative is zero, we can use the third derivative to measure the width of the region where the second derivative remains negligible (i.e., the width of the region where a straight line fit might be valid). We thus propose to define the width $\Sigma_\alpha$ of the transient/anomalous regime (in this specific log-log space) using the expression
\begin{equation}
\label{eq:sigma_alpha}
    \Sigma^3_\alpha = \frac{f(x)}{\sfrac{\partial^3f(x)}{\partial x^3}}\bigg\rvert_{x=log(t_I)}~,
\end{equation}
where $f \! = \! \log (\langle r(t)^2\rangle / 4t )$ , $x \! = \! \log (t)$, and $t_I$ corresponds to the inflection point. The resulting Cartesian temporal width, $\Sigma_t$, is then
\begin{equation}
\label{eq:sigma_t}
    \Sigma_t =  10^{\,[log(t_I)+\Sigma_\alpha]} ~ - ~ 10^{\,[log(t_I)-\Sigma_\alpha]}.
\end{equation}
In principle, we should expect the transient (or anomalous) regime to transition into the steady-state regime when the time $t$ exceeds $\approx \Sigma_t$. Furthermore, we expect this width to diverge at the percolation threshold $\phi^*$ because there is no steady-state at that critical point \cite{King2009}.

\begin{figure}[ht]
    \centering
    \includegraphics[scale=1.0]{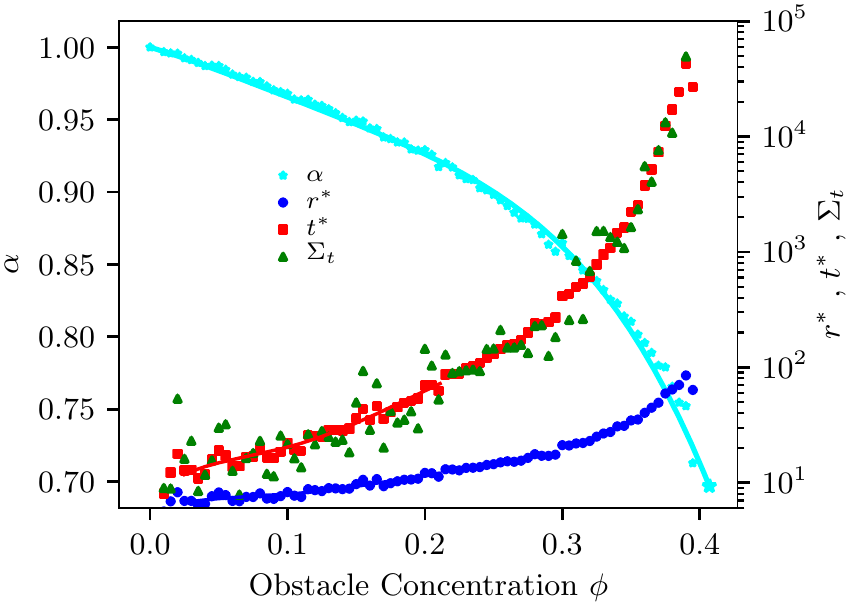}
    \caption{Key system parameters \textit{vs.} obstacle concentration $\phi$; both the obstacles and the particles are of size $1 \times 1$. The crossover length $r^*$ and crossover time $t^*$ diverge as $\phi$ approaches the percolation threshold\cite{jacobsen2015critical} $\phi^* = 0.4072539492079$. The anomalous exponent $\alpha$ decreases (from unity) towards the percolation threshold value\cite{Random_Obstacles_Saxton} $\alpha(\phi \to \phi^*) = \frac{2}{2.73}$. Our estimate of the width of the anomalous regime, $\Sigma_t$, provides an excellent approximation for $t^*$.}
    \label{fig:No_D}
\end{figure}

Figure~\ref{fig:No_D} shows that the width $\Sigma_t$ defined in eq.~\ref{eq:sigma_t} does indeed behave as expected: it increases with $\phi$ and diverges at percolation. Surprisingly, $\Sigma_t$ is found to be essentially equal to $t^*$ over the whole range of concentrations. The fact that it relies on a third derivative of discrete data explains the presence of noise. Unlike $t^*$, the width $\Sigma_t$ does not require any knowledge of the steady-state diffusion coefficient $D$; in other words, it can be obtained using short-time data only.

Since $t^* \approx \Sigma_t$, we can rewrite eq.~\ref{eq:D_Estimate_tStar} as 
\begin{equation}
    \label{eq:Destimate}
    D \approx D_\alpha \, \Sigma_t^{\alpha-1}~.
\end{equation}
We stress again the fact that the three parameters on the \textit{rhs} of this expression can be obtained using transient data only. Therefore, eq.~\ref{eq:Destimate} implies that it is possible to obtain an estimate of the steady-state diffusion coefficient $D$ that uses only parameters extracted from the transient regime (there is no need to reach the steady-state). We test this approach to estimating $D$ in the next two sections for several sizes of random-walkers and obstacles.

\subsection{Estimating the diffusion coefficient $D$ using $\Sigma_t$}
\label{sec:testing}

\begin{figure}[ht]
    \centering
        \includegraphics[scale=1]{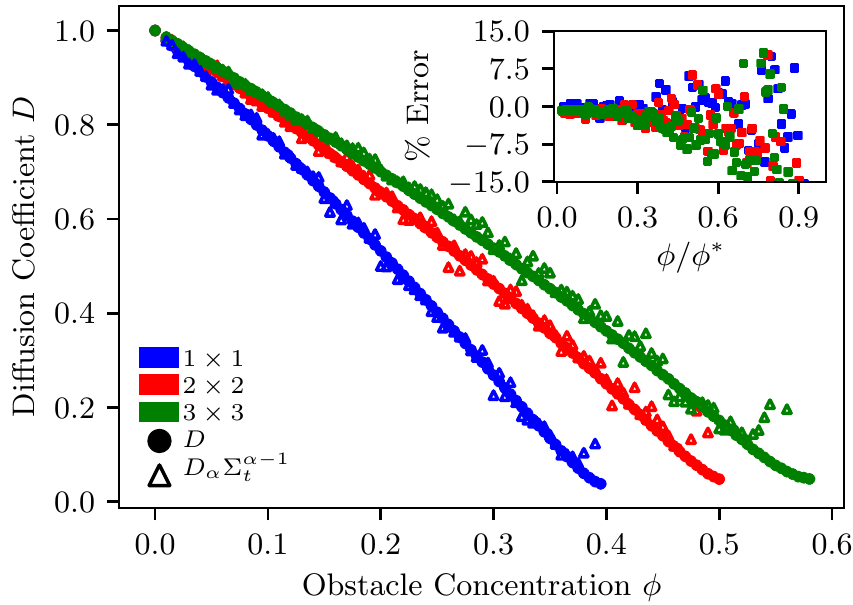}
\caption{
Diffusion coefficient of a $1 \times 1$ particle \textit{vs} obstacle concentration $\phi$ (fraction of the surface area covered by obstacles). The solid circles correspond to the high-precision values of $D$ obtained using the matrix method; the triangles give the approximation $D_{\alpha}\Sigma_t^{\alpha-1}$ as defined by eq.~\ref{eq:Destimate}. Results are shown for three different obstacle sizes. Inset: relative error $\left[\frac{ D - D_{\alpha}\Sigma_t^{\alpha-1}}{D}\right]$ (in \%) \textit{vs} obstacle concentration divided by the percolation threshold $\phi^*$.}
    \label{fig:D}
\end{figure}

\begin{figure}[ht]
    \centering
    \includegraphics[scale=1]{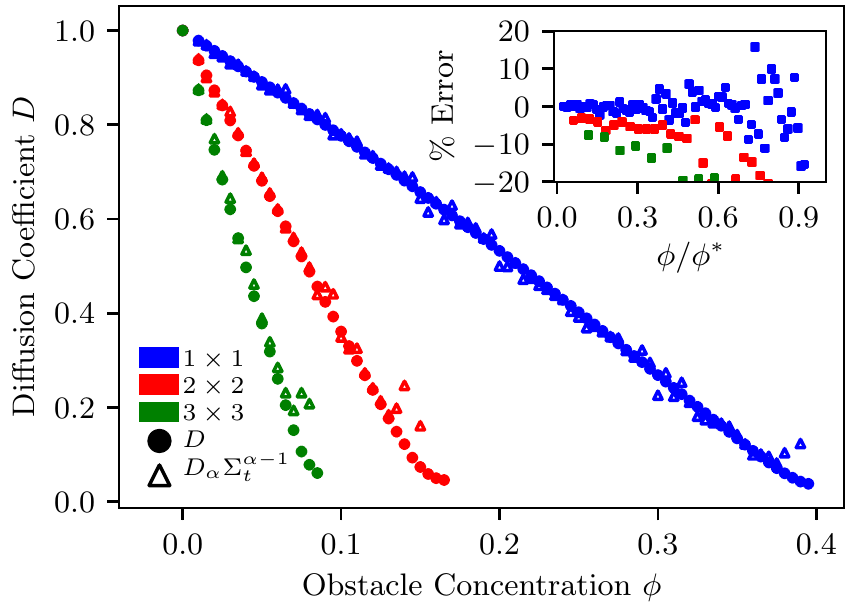}
    \caption{Diffusion coefficient \textit{vs} concentration of $1 \times 1$ obstacles. The solid circles correspond to the high-precision values of $D$ obtained using the matrix method; the triangles give the approximation $D_{\alpha}\Sigma_t^{\alpha-1}$ as defined by eq.~\ref{eq:Destimate}. Results are shown for three different random walker sizes. Inset: relative error $\left[\frac{ D - D_{\alpha}\Sigma_t^{\alpha-1}}{D}\right]$ (in \%) \textit{vs} obstacle concentration divided by the percolation threshold $\phi^*$.}
    \label{fig:D_Walk}
\end{figure}

In order to test the accuracy of eq.~\ref{eq:Destimate}, we compare its prediction to exact values obtained using the method described in Section~\ref{Methods}, and we do this for six different systems of random walkers and randomly distributed obstacles in Figs.~\ref{fig:D} (for obstacles of different sizes) and~\ref{fig:D_Walk} (for random walkers of different sizes).

As Fig.~\ref{fig:D} shows, the approximation holds very well for $1 \times 1$ particles and different obstacle sizes across all obstacle concentrations. Figure~\ref{fig:D_Walk} shows equally good results when larger particles move between small obstacles. In both cases, the results become more noisy and less reliable near the relevant percolation threshold $\phi^*$ because of finite size effects. Indeed, the last few points of the exact diffusion constant data set in Figs.~\ref{fig:D} and~\ref{fig:D_Walk} show a change in curvature; this is the result of the crossover length $r^*$ increasing quickly near percolation (see Fig.~\ref{fig:No_D}), and finite size effects becoming prominent (as the condition $L \gg r^*$ is no longer valid). 
\begin{table*}[t]
    \begin{tabular}{|c|c|c|c|}
        \hline 
         & $1 \times 1$~ (see~ ref.~\cite{jacobsen2015critical}) & $2 \times 2$ & $3 \times 3$ \\
         \hline
         Obstacles & 0.40725394920790(2) & 0.51(1) & 0.58(1) \\
         Walkers & 0.40725394920790(2) & 0.161(4) & 0.083(1) \\
         \hline
    \end{tabular}
    \caption{Percolation threshold $\phi^*$ for walkers and obstacles of different sizes. When investigating different sizes of obstacles or walkers the counterpart remains at a size of $1 \times 1$.}
    \label{tab:PercThresh}
\end{table*}

To estimate the percolation thresholds for our systems, we performed a linear fit on the final 10 data points prior to the change of curvature observed due to finite size effects. This linear fit is extrapolated to $D\to0$ to estimate $\phi^*$. The results are given in Table~\ref{tab:PercThresh}. The insets in Figs.~\ref{fig:D} and~\ref{fig:D_Walk} show the relative error in our estimates of $D$ as a function of the scaled concentration $\phi/\phi^*$. These errors increase with the concentration and become of the order 15\% as we approach the percolation thresholds here; as usual, these errors decrease with the ensemble size (data not shown). Note that the three curves in the main parts of Figs.~\ref{fig:D} and \ref{fig:D_Walk} nearly collapse on a universal curve if $D$ is plotted as a function of $\phi/\phi^*$, as one would expect (data not shown).

\section{Conclusion}

In this paper, we first proposed a non-arbitrary way to characterize the transient/anomalous regime in the case of obstructed diffusion problems. In particular, we introduced a method to locate the center of this regime and estimate the value of the related anomalous exponent $\alpha$. Furthermore, we suggested a way to measure the width $\Sigma_t$ of the regime during which the MSD could potentially be fit with $\left< r^2 (t) \right> \sim t^\alpha$.  

Using simple two-dimensional lattice models of obstructed diffusion, we found that the width $\Sigma_t$ of the anomalous regime can act remarkably well as a proxy for the crossover time $t^*$ marking the transition between the anomalous and steady-state regimes. For instance, both diverge identically near the percolation threshold. This allows us to write eq.~\ref{eq:Destimate} which yields an estimate of the steady state diffusion coefficient $D$ using only short time data. Our simulations have shown that this method for estimating the steady state diffusion coefficient is robust for a variety of random obstacle variants (i.e., larger obstacles, and larger walkers). 

One drawback to this approach is the need to evaluate a third-derivative to compute $\Sigma_t$ and hence $D$. This may limit the usefulness of our findings when the data are noisy (we saw examples of this when close to percolation thresholds). Clearly, this needs to be explored further, e.g. using data coming from Molecular Dynamics simulations or experimental data.

Experimentally (and with computer simulations), our novel data analysis method can allow one to access the steady-state diffusion coefficient where otherwise it would be inaccessible due to the inability to reach late time data. Interestingly, eq.~\ref{eq:r*} can also be rewritten as $r^{*^2}=4D_\alpha \Sigma_t^\alpha$; this means that the correlation length $r^{*^2}$ can also be estimated using short-time data. 

\begin{acknowledgments}
NI thanks the University of Ottawa for an admission scholarship. GWS acknowledges the support of both the University of Ottawa and the Natural Sciences and Engineering
Research Council of Canada (NSERC), funding reference number RGPIN/046434-2013.
\end{acknowledgments}

\bibliography{References}

\end{document}